\documentclass[11pt,twoside]{article}

\usepackage{asp2004}
\usepackage{epsf}
\usepackage{psfig}
\usepackage{epsfig}
\usepackage{lscape}
\usepackage{times}

\begin{document}
\title{3D photoionisation and dust RT modelling with MOCASSIN: geometry effects on the emission line spectra of star-forming regions}    
\author{Barbara Ercolano \& Nate Bastian}   
\affil{Dept of Physics and Astronomy, University College London, UK}    

\begin{abstract} 
 Emission line spectra from H~{\sc ii} regions are often used to study properties of the gas in star-forming regions, as well as temperatures and luminosities of the ionising sources. Empirical diagnostics for the interpretation of observational data must often be calibrated with the aid of photoionisation models. Most studies so far have been carried out by assuming spherical or plane-parallel geometries, with major limitations on allowed gas and dust density distributions and with the spatial distribution of multiple, non-centrally-located ionising sources not being accounted for. We present the first results of our theoretical study of geometric effects, via the construction of a number of 3D photoionisation models using the {\sc mocassin} code for a variety of spatial configurations and ionisation sources. We compare integrated emission line spectra from such configurations and show evidence of systematic errors caused by the simplifying assumption of a single, central location for all ionising sources. 
\end{abstract}


\section*{Effects of 3D geometry on the emission line spectra of ionised regions}

\paragraph{The {\sc mocassin} code:} Spatially-resolved studies of star-forming regions indicate that the assumption of spherical geometry is not realistic in most cases, further complication is added by the gas being ionised by multiple non-centrally located sources. 
The 3D photoionisation code {\sc mocassin} was designed to treat all of the above, allowing for the transfer of both primary and secondary radiation to be treated self-consistently, without the need of approximations. The code was thoroughly benchmarked \citep{code01,code02} and has been applied to the study of several ionised regions. The current version includes a fully self-consistent treatment of the radiative transfer of dust grains mixed within the gas, taking into account the microphysics of dust-gas interactions within the geometry-independent Monte Carlo transfer. 

\paragraph{3D gas and star distribution on the emission line spectra of ionised regions:} We are investigating geometric effects theoretically, via the construction of a number of 3D photoionisation models for a variety of spatial configurations and ionisation sources. We compare integrated emission line spectra from such configurations in search of systematic errors, which may be caused by the simplifying assumption of a single central location for all ionising sources. 
Our study is initially tailored at H~{\sc ii} region spectra, but will successively be extended to star clusters and cluster complexes.

\begin{figure*}
\begin{center}
\begin{minipage}[t]{6.5cm}
\includegraphics[width=6.5cm]{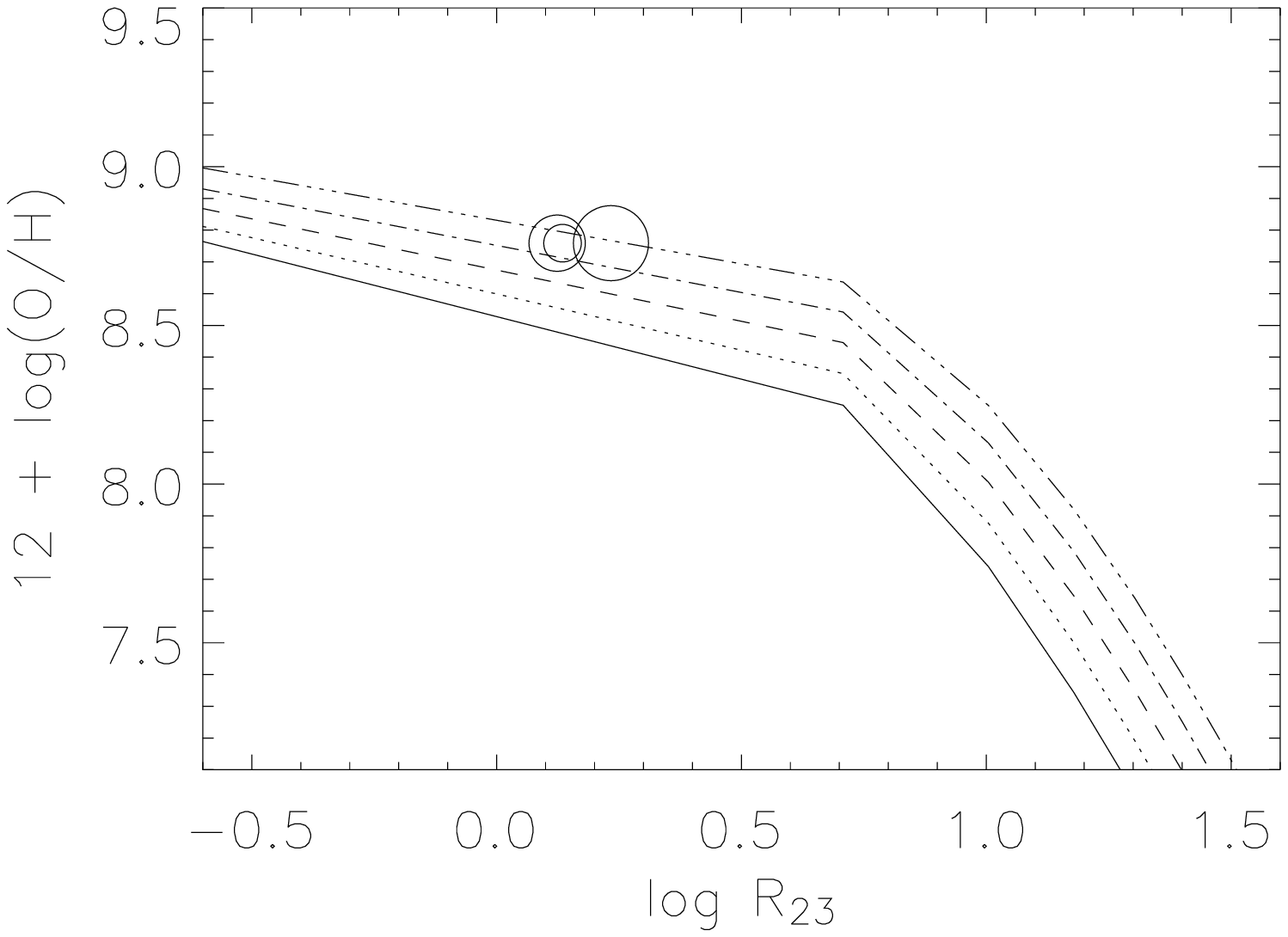}
\end{minipage}
\begin{minipage}[t]{6.5cm}
\includegraphics[width=6.5cm]{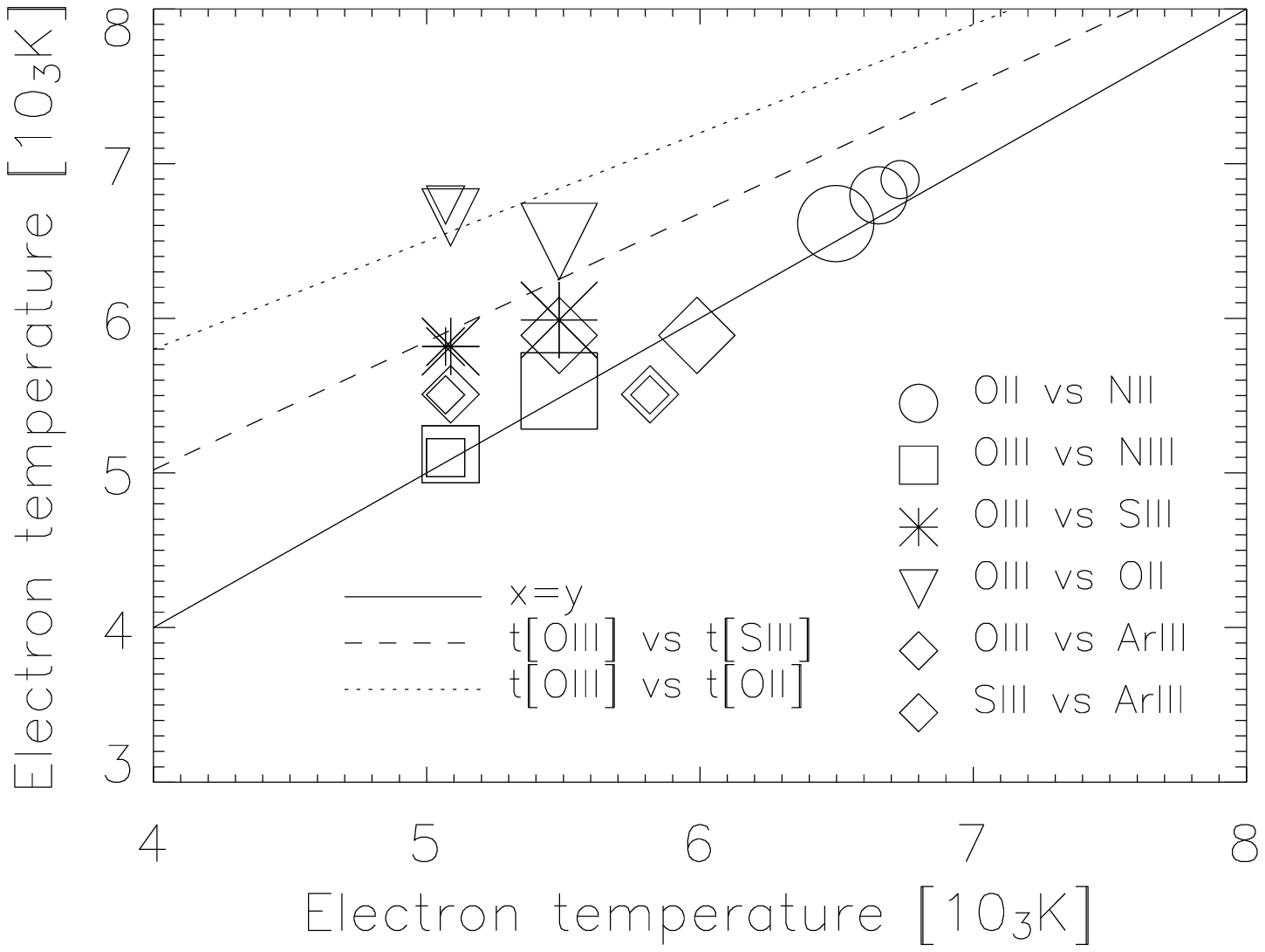}
\end{minipage}
\caption[]{Homogeneous sphere model; small, medium and large circles (P=0.72, 0.74, 0.62) for central, half- and full-volume distributed sources. {\it Left:} Model R23 ratios (circles) compared to \cite{code04} calibrations for P=0.2 (solid line), 0.35, 0.5, 0.65 and 0.80. {\it Right:} Model ionic temperature (in [kK]) compared to assumed relations; solid line is for equal temperatures, dotted and dashed lines are the t$[$O~{\sc iii}$]$, t$[$O~{\sc ii}$]$ and t$[$O~{\sc iii}$]$, t$[$S~{\sc iii}$]$ relations of \cite{code05}.}
\end{center}
\end{figure*}

Our preliminary models are based on homogeneous or porous spherical density distributions ionised by 240 sources, comprising of 37\,M$_{\odot}$ (35 kK) stars and 56\,M$_{\odot}$ (56 kK) stars. The stars are either located at the centre or distributed in the half- or full-spherical volume.
The ionising fluxes were calculated using {\sc starburst99} \citep{code03}, assuming solar metallicities and an age of 1Myr. 
\paragraph{\bf Preliminary results: R23 O-abundance diagnostic}
The sum of the bright O forbidden lines ($[$O~{\sc iii}$]\lambda\lambda$5007,4959 + $[$O~{\sc ii}$]\lambda\lambda$3727,29)/H$\beta$ is often the sole handle on metallicity for extragalactic H~{\sc ii} regions and starburst galaxies. 
We tested the robustness of abundances derived via this method using the calibration of \cite{code04}. O abundance in our models for central half-volume distributed sources agree extremely well to the values obtained using \cite{code04} calibrations for the appropriate excitation index, $P$. Significant deviations are shown, however, by the models with full-volume distributed sources For an homogeneous sphere ($n_{\rm H}$=100\,cm$^{-3}$), this implies that {\it O abundances could be underestimated by $\sim$25\% by empirical methods} (Fig~1, left). Similar results were found for the other density distributions investigated so far. 

\paragraph{\bf Preliminary results: Ionic temperature relations}
We tested the robustness of relations commonly assumed to hold between different ionic temperatures. Good agreement is found for central and half-volume stellar configuration cases in homogeneous density spheres, however models where the stars are distributed over the full volume have significantly lower temperatures than would be estimated from 1D models (Fig.~1, left) for lower excitation species, such as O$^+$ and N$^+$ and higher for higher excitation ones. This means that the t$[$O~{\sc iii}$]$, t$[$O~{\sc ii}$]$ relation of \cite{code05} fails for the full-volume stellar distribution models. Similar results were found for porous density distributions; here, however, larger variations are found for the t$[$O~{\sc iii}$]$, t$[$O~{\sc ii}$]$ relation, with t$[$O~{\sc ii}$]$ severely underestimated for all stellar configurations.




\end{document}